\documentclass[aps,prl,twocolumn,groupedaddress]{revtex4}  
\usepackage{graphicx}  
\usepackage{dcolumn}   
\usepackage{bm}        
\usepackage{amssymb}   

\newcommand{\eqref}[1]{{(\ref{#1})}}

\newcommand{\be}{\begin{equation}}
\newcommand{\ee}{\end{equation}}
\newcommand{\bea}{\begin{eqnarray}}
\newcommand{\eea}{\end{eqnarray}}
\newcommand{\nn}{\nonumber}

\newcommand{\ti}{\times}

\newcommand{\mc}{\mathcal}

\newcommand{\beqa}{\begin{eqnarray}}
\newcommand{\eeqa}{\end{eqnarray}}

\begin{document}


\title{Searching for a $0.1-1$ keV Cosmic Axion Background}
\author{Joseph P.~Conlon}\email{j.conlon1@physics.ox.ac.uk} \affiliation{Rudolph Peierls Centre for Theoretical Physics, University of Oxford, 1 Keble Road, OX1 3NP, Oxford, United Kingdom}
\author{M.C.~David Marsh}\email{david.marsh1@physics.ox.ac.uk}\affiliation{Rudolph Peierls Centre for Theoretical Physics, University of Oxford, 1 Keble Road, OX1 3NP, Oxford, United Kingdom}
\date{\today}

\begin{abstract}
Primordial decays of string theory moduli at $z \sim 10^{12}$  naturally generate a dark radiation Cosmic Axion Background (CAB) with 0.1 - 1 keV energies. This CAB can be detected through axion-photon conversion in astrophysical magnetic fields to give
quasi-thermal excesses in the extreme ultraviolet and soft X-ray bands. Substantial and observable luminosities may be generated even for axion-photon couplings
$\ll 10^{-11} \hbox{GeV}^{-1}$. We propose that axion-photon conversion may explain the observed excess emission of soft X-rays from galaxy clusters,  and may also contribute to the diffuse unresolved cosmic X-ray background. We  list a number of correlated predictions of the scenario.


\end{abstract}

\maketitle


The success of the simple $\Lambda$CDM model cannot obscure the fact that it will not be the last word in cosmology. One natural
extension of $\Lambda$CDM is to include  an extra, relativistic contribution to the energy density 
of the universe. Such \emph{dark radiation}
is  conventionally
parametrised as an excess number of neutrino species, $\Delta N_{eff} = N_{eff} - 3.046$. There are indeed observational hints for such a contribution: including the HST measurement of the Hubble constant \cite{11032976}, the current observational values from WMAP, ACT, SPT and Planck are $N_{eff} = 3.84 \pm 0.40$ (WMAP9, \cite{12125226}), $3.71 \pm 0.35$ (SPT, \cite{12126267}), $3.50 \pm 0.42$ (ACT, \cite{13010824}),  and $3.62 \pm 0.25$ (Planck, \cite{13035076}). Without combining with $H_0$, the values for $N_{eff}$ are $N_{eff} = 3.55 \pm 0.60$ (WMAP9), $2.87 \pm 0.60$ (ACT), $3.50 \pm 0.47$ (SPT) and $3.30 \pm 0.27$ (Planck).

Dark radiation is a theoretically intriguing  extension to $\Lambda$CDM as it is a natural consequence of simple and appealing models of the early universe. The standard post-inflationary picture of the early universe involves the reheating of the Standard Model (SM) from the decays of a scalar field. In addition to its decay modes to visible sector SM particles, this field may also decay to (effectively) massless weakly coupled hidden sectors, such as axions or hidden photons. If these particles are sufficiently
weakly coupled to SM matter, they will not thermalise and will remain as relativistic dark radiation, redshifting with the
expansion of the universe.

This picture is particularly well motivated within string theory models of the early universe. As radiation redshifts as $a^{-4}$ and matter redshifts as $a^{-3}$, we expect reheating to be driven by the last scalar field to decay. String theory contains many light scalar fields, called moduli, with gravitational strength couplings. Such moduli are very long-lived, with life-times
\be
\tau \sim 8 \pi \frac{M_{Pl}^2 }{m_{\Phi}^3} \, ,
\ee
where $M_{Pl} = 2.4 \ti 10^{18}~\hbox{GeV}$.
Moduli generically couple both to visible SM matter and any hidden axions that are present \cite{12083562, 12083563, 13047987}.

As the energy density at decay is $V = 3 H_{decay}^2 M_{Pl}^2$ with $H_{decay} \sim \tau^{-1}$, the SM reheat
temperature is
\be
T_{reheat} \sim \frac{m_{\Phi}^{3/2}}{M_{Pl}^{1/2}} \sim \hbox{0.6 GeV} \left( \frac{m_{\Phi}}{10^6 \hbox{GeV}} \right)^{3/2} \, .
\ee
Hidden sector decays $\Phi \to aa$ generate a Cosmic Axion Background (CAB). As these decays are 2-body with $E_a = m_{\Phi}/2$, the CAB
energies are substantially greater than the SM reheat temperature, by a factor $\left( M_{Pl}/m_{\Phi} \right)^{1/2}$.
\begin{figure}
\includegraphics[scale=0.45]{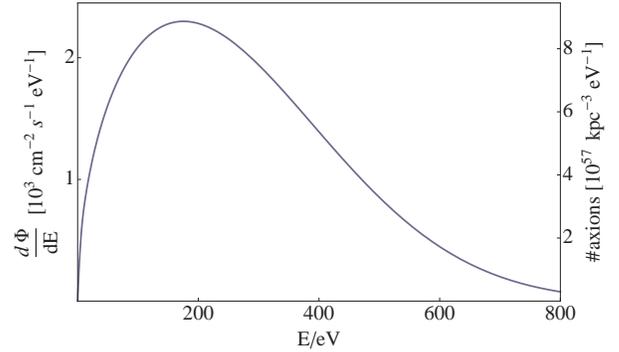}
\caption{\label{fig:spectrum} The present day axion spectrum resulting from the decay of a modulus of mass $m_{\Phi} \sim 10^6$ GeV. 
}
\end{figure}

This ratio is effectively maintained throughout cosmic history, up to small $g_{*}^{1/3}$ boosts in the photon temperature,
and therefore also sets the present day axion energies relative to the CMB.
In these expressions the moduli mass $m_{\Phi}$ is unspecified, although the requirement that reheating occurs before BBN implies
$m_{\Phi} \gtrsim 30~\hbox{TeV}$. The moduli masses are normally comparable to or slightly larger than the scale of supersymmetry breaking.
If supersymmetry is relevant to the  hierarchy between the Planck scale and the weak scale, the lightest modulus is expected to have $m_{\Phi} \lesssim 10^7 \hbox{GeV}$
and in many models has $m_{\Phi} \sim 10^6~\hbox{GeV}$ (cf.~\cite{09063297} for the LARGE volume scenario, and
\cite{hepph0503216, 08040863}
for other work).
In \cite{13041804}, we pointed out that this gives rise to a prediction of a CAB with ${\cal O}(E) \sim 200$ eV and a homogeneous and isotropic flux of $10^6$ cm$^{-2}$ s$^{-1}$. The (non-thermal) spectral shape of the CAB arises from  modulus decays and --- as is shown in figure \ref{fig:spectrum} ---   has a `quasi-thermal'
shape  \cite{13041804}.

This CAB would have freely propagated since $z \sim 10^{12}$ ($t \sim 10^{-6}~\hbox{s}$), which is a factor of $10^{19}$ earlier in time than the CMB. This would provide a
spectacular probe of the early universe.
In this Letter, we search for signatures of this CAB through $a \to \gamma$ conversion in astrophysical and cosmological magnetic fields.

Although subdominant in energy density to either baryonic matter or the CMB, the
energy density associated to the CAB is still substantial,
\be
\rho_{CAB} = 1.6 \cdot 10^{60}~\hbox{erg}~\hbox{Mpc}^{-3}  \left( \frac{\Delta N_{eff}}{0.57} \right)\, ,
\ee
and entirely located in the extreme ultraviolet (EUV) and soft X-ray bands. A galaxy cluster occupies an approximate volume of  $1 \hbox{ Mpc}^3$ and has a typical X-ray luminosity of $\mc{L}_{cluster} \sim 10^{44}~\hbox{erg} \, \hbox{s}^{-1}$. This makes it clear that even a very small $a \to \gamma$ conversion rate will generate a large signal.

Axion-photon conversion is well known to occur in the presence of coherent magnetic fields \cite{Sikivie1, Sikivie2}.
The axion-photon Lagrangian is given by,
\bea
{\cal L} &=& -\frac{1}{4} F_{\mu \nu}F^{\mu \nu} - \frac{1}{4 M}  a F_{\mu \nu} \tilde F^{\mu \nu} \nonumber  \\
 &+& \frac{1}{2} \partial_{\mu}a \partial^{\mu}a - \frac{1}{2} m_a^2 a^2 \, ,
\eea
where the coupling $M^{-1}$ has dimension $-1$ and gives rise to oscillations between axions and photons. Here, the axion field $a$ is a general pseudo Nambu-Goldstone boson of a broken shift symmetry which need not correspond to the QCD axion \cite{Peccei:1977hh}.

For this  case $M$ and $m_a$ are uncorrelated and $a$ is sometimes called an axion-like particle. We will mostly be interested in $m_a \lesssim 10^{-9}$ eV,
where direct bounds are $M \gtrsim 10^{11}$ GeV.

The $a \to \gamma$ conversion probability for an axion in a coherent magnetic field domain of length $L$ and with transverse component $B_{\perp}$ can be computed by elementary methods  \cite{Raffelt:1987im}, and is given by,
\be
\label{def}
P(a \rightarrow \gamma) = \sin^2(2 \theta) \sin^2\left( \frac{\Delta}{\cos 2 \theta}\right) \, , \label{P}
\ee
where  $\tan 2 \theta = \frac{2 B_{\perp} \omega}{M m^2_{eff}}$, $\Delta = \frac{m_{eff}^2 L}{4 \omega}$,  $m^2_{eff} = m_a^2 - \omega_{pl}^2$,  $\omega_{pl}$ is the plasma frequency,
\bea
\omega_{pl} = \left(4 \pi \alpha \frac{n_e}{m_e}\right)^{1/2} =1.2\cdot10^{-12} \sqrt{\frac{n_e}{10^{-3} {\rm cm}^{-3}}}~{\rm eV} \nonumber \, ,
\eea
and $\omega$ denotes the photon energy. Though not crucial for our analysis, we note that in the small-angle approximation $\theta \ll1$ and $\Delta \ll1$, the conversion probability is simply given by
\be
\label{abc}
P(a \rightarrow \gamma) = \frac{1}{4} \left( \frac{B_{\perp} L}{M}\right)^2 \, . \label{Papprox}
\ee
We will apply equations (\ref{def}) and (\ref{abc}) to obtain signatures of the CAB. To allow an easy estimation of magnitudes,
we will generally quote results within the small angle approximation, but for plots we shall use the full expression in equation (\ref{def}).



Axion-photon conversion is maximised in regions of large coherent magnetic fields. Galaxy clusters and
the intracluster medium (ICM) provide such regions. The existence of magnetic fields in the ICM has been established by a number of methods, with typical values of ${\cal O }(B) \sim \mu \hbox{G}$, and with larger values observed close to cluster cooling cores \cite{Govoni:2004as}. The coherence lengths of these fields are not known in detail, but are expected to be in the range 
$L \sim 1-10$ kpc.

The CAB axions  will convert into photons with energies ${\cal O}(\omega) = 0.1 - 1$ keV, and for very small axion masses where $m^2_{eff} = \omega_{pl}^2$ 
 we  find,
\bea
\theta &\approx& \frac{B_{\perp} \omega}{M m^2_{eff}} = 2.8 \cdot 10^{-5} \times \left( \frac{10^{-3} \hbox{cm}^{-3}}{n_e} \right) \nonumber \\
&\times&  \left(\frac{B_{\perp}}{1~{\rm \mu G} }\right)  \left(\frac{\omega}{200~{\rm eV} }\right)  \left(\frac{10^{14}~{\rm GeV} }{M }\right) \, . \label{eq:theta}
\eea
The small-$\theta$ approximation is then almost always justified.
We also have
\be
\Delta = 0.27 \times  \left( \frac{n_e}{10^{-3} \hbox{cm}^{-3}} \right) \left(\frac{200~{\rm eV}}{\omega} \right) \left(\frac{L}{1~{\rm kpc}}\right)\, . \label{eq:Delta}
\ee
The small-$\Delta$ approximation is then only valid for a limited parameter range.

For illustration, we first work in a parameter regime where
 \eqref{Papprox} applies.
Axions travelling through a 1 kpc ICM magnetic field convert to photons with probability
\be
P(a \rightarrow \gamma) = 2.3 \cdot 10^{-10}\times \left(\frac{B_{\perp}}{1~{\rm \mu G} }\frac{L}{1~{\rm kpc}}\frac{10^{14}~{\rm GeV} }{M }\right)^2 \, .
\ee
The corresponding conversion rate per axion per second is $2.3 \cdot 10^{-21}$ s$^{-1} \times \left(\frac{B_{\perp}}{1~{\rm \mu G} }\frac{10^{14}~{\rm GeV} }{M }\right)^2 \left(\frac{L}{1~{\rm kpc}}\right)$.

We now for simplicity assume that the dark radiation is dominated by a single axion species and compute the induced luminosity from $a \rightarrow \gamma$ conversion. As the axion flux is homogeneous and isotropic, we average over the alignment of the axion velocity to the magnetic field, giving $\langle B_{\perp}^2\rangle = \frac{1}{2} B^2$, where $B$ denotes the magnitude of the magnetic field. Summing over magnetic domains, the luminosity from $a \to \gamma$ conversion is
\bea
{\cal L}_{Mpc^3} 
&=& 3.6 \cdot 10^{39}~\hbox{erg Mpc}^{-3} \hbox{s}^{-1} \times  \label{flux} \\
&\times&\left(\frac{\Delta N_{eff}}{0.57} \right)\left(\frac{B}{\sqrt{2}~{\rm \mu G} }    \frac{10^{14}~{\rm GeV} }{M }\right)^2 \left( \frac{L}{1~{\rm kpc}} \right)
\, , \nonumber
\eea
and lies dominantly in the EUV and soft X-ray bands.
In the small angle approximation, the shape of the resulting photon spectrum is identical to the axion spectrum in figure \ref{fig:spectrum}. Beyond this approximation the conversion probability is dependent on $\omega$, and the resulting photon spectrum is obtained by weighting the axion spectrum with equation \eqref{P}.

\begin{figure}
\includegraphics[scale=0.45]{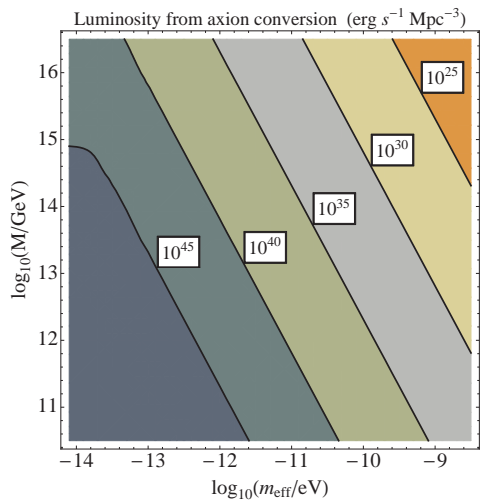}
\caption{\label{fig:flux} The CAB luminosity from axion conversion with $\Delta N_{eff} = 0.57$, $B_{\perp} = 1~{\rm \mu G}$ and $\omega=200~{\rm eV}$. The distribution of   $L$ is taken to be uniform in the range of $2-34$ kpc and is marginalised over.
}
\end{figure}


Since the launch of EUVE (Extreme Ultraviolet Explorer), excess EUV and soft X-ray cluster emission above the hot ($T \sim 5\div10~\hbox{keV}$) intracluster medium has been observed  by all major space telescopes with soft X-ray sensitivity in a large number of galaxy clusters. As is reviewed in \cite{Durret:2008jn}, these include
EUVE \cite{Lieu1996}, ROSAT \cite{Bonamente2002}, BeppoSAX \cite{Bonamente2001}, XMM-Newton
\cite{Nevalainen2003}, Chandra \cite{Henriksen2003, Bonamente2007} and Suzaku \cite{Lehto2010}.

One particularly well studied cluster is  Coma, which is large, luminous and nearby.
The soft X-ray excess has been well-documented in Coma since the original discovery \cite{Lieu1996}. It has been established that the soft X-ray excess is diffuse and extends beyond the region containing the hot intracluster gas,
up to $5$ Mpc from the cluster centre \cite{Bonamente:2009ns}. Based on data in \cite{Lieu1996}, the excess luminosity in the $0.1-1$ keV range within a central 18 arcminute radius (corresponding to $r\leq 0.50$ Mpc with $H_0 = 73$ km s$^{-1}$ Mpc$^{-1}$) of the cluster centre is $1.6 \cdot 10^{42}$ erg/s. Within radii of 1 Mpc, the magnetic field strength in the Coma cluster has been measured  to be around $2-5~\mu$G with coherence lengths ranging from $2$ kpc to $34$ kpc \cite{10020594}.

Proposed astrophysical explanations of the soft excess include either thermal emission from a warm $T \sim 0.2~\hbox{keV}$ gas or
non-thermal inverse Compton scattering of relativistic electrons
off CMB photons (IC-CMB, e.g. \cite{Ensslin1999, Bowyer2004}).
The former explanation is problematic for two reasons. First, a thermal gas should also generate emission lines (for example from $\hbox{O}_{\hbox{VII}}$ at 560 \hbox{eV}), and no such lines have been observed. Secondly, such a gas would cool very rapidly as
it requires a large density to maintain pressure against the intracluster medium.

The ostensible explanation of the soft excess by IC-CMB 
is also problematic.
By construction, it requires a large population
of relativistic electrons with $\gamma \sim 300$ that scatter off the CMB. As discussed 
 in \cite{AtoyanVolk, TsayBowyer}, this population is constrained by its synchrotron emission in radio frequencies and its bremmstrahlung emission in gamma rays.

 Most proposed models of IC-CMB assume either a power-law shape of the electron number density density from $\gamma \sim 300$ (where it is fixed by the observed soft excess) to $\gamma \sim 3000$ (where it emits radio-frequency synchrotron radiation), or a power-law with a spectral break at some intermediate value of $\gamma$.

 The limits on
  radio emission from the Coma radio halo together with improved determinations of the Coma magnetic field \cite{10020594} place stringent constraints on the IC-CMB models: since the synchrotron emission scales as $B^2$, models which were plausible for magnetic field values of $0.2 \div 0.5~\mu$G are immediately excluded for $2 \div 5~\mu$G Coma magnetic fields.  We note that these improved observations exclude the vast majority of all models of the soft excess as arising from IC-CMB.


Any viable IC-CMB model must therefore have a sharp spectral cutoff between $\gamma \sim 300$ and $\gamma \sim 3000$. 
This cutoff may potentially be generated from a single injection event at some point in the past, followed by subsequent radiative losses. However, independently of the spectral shape at large $\gamma$, the electrons with $\gamma \sim 300$ (whose number density, by assumption, is fixed by the magnitude of the soft excess) must still emit gamma rays through bremmstrahlung. In \cite{AtoyanVolk} this emission was calculated and found to be well in excess of the FERMI-LAT sensitivity. As Fermi has not observed any clusters in gamma rays \cite{Han, Ando}, such IC-CMB models appear to be ruled out.

Thus,  recent observations appear to have ruled out the proposed astrophysical models for the soft excess in Coma, either through
the combination of the comparatively large 
 magnetic field and the limits on radio emission, or through the failure of Fermi-LAT to observe clusters in gamma rays. While it would be premature to conclude that no astrophysical model can work, this motivates consideration of alternative explanations.

From equation \eqref{flux}, we note that in the small angle approximation, the luminosity from axion-photon conversion in a cylindrical volume with radius $0.5$ Mpc and length  $3$ Mpc is given by
\bea
{\cal L}
&=& 1.7\cdot10^{42}~ \hbox{erg s}^{-1} \times \\
&\times&\left(\frac{\Delta N_{eff}}{0.57} \right)\left(\frac{B}{2~{\rm \mu G} }\frac{10^{13}~{\rm GeV} }{M }\right)^2 \left( \frac{L}{1~{\rm kpc}} \right)
\, . \nonumber
\eea
In figure \ref{fig:flux}, we show the total luminosity per Mpc$^3$,  where we have also marginalised over $L$. This plot is evaluated using the
full expression eq. (\ref{def}) and is not restricted to the small angle approximation.

As the direct constraint on $M$ for light axions is only $M \gtrsim 10^{11} \hbox{GeV}$
we conclude that axion-photon conversion in the Coma ICM may easily give rise to a soft X-ray excess of the observed order of magnitude.

In addition to reproducing the soft X-ray excess, our model makes several additional predictions.
As the CAB is uniformly distributed, the produced luminosity is determined only by the magnetic field and the electron density, and is independent of the cluster temperature or matter
distribution. The model predicts that the soft excess should be largest in cluster regions with large magnetic fields and small electron densities,
and its spatial extent should be coterminous with the magnetic field. Magnetic field strengths of the order of $0.5~\mu$G have measured
in the `bridge' region of the Coma-Abell 1367 supercluster, at a radius of $\sim 1.5~\hbox{Mpc}$ from the central region of the Coma cluster  \cite{Kim1989} . Thus, our model appears consistent with the large radial extent of the Coma soft excess emission.

The formula eq. (\ref{def}) is probabilistic, and could therefore fluctuate depending on the details of the actual realisation of the magnetic field
(discussed for example in \cite{09110015}). This could have a significant effect on observations of a point source. However for the case of clusters the effects of such stochastic fluctuation should be softened when performing an angular average of the soft excess at a fixed radius from the cluster center.

In this model X-ray photons arise non-thermally from $a\rightarrow\gamma$ conversion. It is therefore a clear prediction
 that it should not be possible to associate any thermal emission lines (e.g from $\hbox{O}_{\hbox{VII}}$ at 561, 569 and 574 eV) to the soft excess.

Furthermore, the CAB axions are redshifting, and used to be more
energetic by $(1+z)$ and more dense by $(1+z)^3$. It is then a prediction that the energy scale of the soft excess should grow as $(1+z)$
and, if other aspects of cluster physics are identical, the overall energy in the soft excess should grow as $(1+z)^4$.

In addition to cluster spectra, $a \to \gamma$ conversion in large-scale cosmological magnetic fields may also contribute to the diffuse unresolved cosmic X-ray background (CXB) in the same 0.1-1 keV band.

In the 0.5-2 keV region the diffuse CXB is $8.2 \cdot 10^{-12}  \hbox{ erg cm}^{-2} \hbox{ s}^{-1} \hbox{ deg}^{-2}$, or in total,
$3.4 \cdot 10^{-7}  \hbox{ erg cm}^{-2} \hbox{ s}^{-1}$. After subtracting Chandra and HST sources, the diffuse CXB is essentially removed in the 1-2 keV band but remains present in the 0.5-1 keV band, suggesting a different and genuinely diffuse origin for the unresolved CXB below 1 keV  \cite{0512542, 0702556, 12084105}. In \cite{0702556} the residual $0.65-1$ keV diffuse intensity was  given as $(1.0 \pm 0.2) \cdot10^{-12}$ erg cm$^{-2}$ s$^{-1}$ deg$^{-2}$.

Here we compute the $a\rightarrow \gamma$ contribution to the CXB
for certain illustrative parameters.  We take  $B_{\perp} =1~\hbox{nG}$ magnetic fields and $L=10$ Mpc scales,  although we caution that the actual magnitude of cosmological magnetic fields is unknown. For an electron density equal to the cosmological baryon density $n_B = 2.5\cdot 10^{-7}$ cm$^{-3}$, and $m_{eff} = \omega_{pl}$, we find that the  $a \to \gamma$ conversion probability  is $2.0 \cdot 10^{-6}$ per coherent domain for $M=10^{13}$ GeV. As a rough approximation, we assume that these conditions have held for $10^{10}$ years, and we average over the direction of the magnetic field to obtain the total conversion probability per axion,
\bea
P(a \to \gamma) = 6.1 \cdot 10^{-4} \left( \frac{B}{\sqrt{2} \hbox{ nG}} \frac{10^{13} \hbox{GeV}}{M} \right)^2. \nn
\eea
As the axion flux on earth  is $4.1 \cdot 10^{-4}~\hbox{erg cm}^{-2} \hbox{ s}^{-1}$, the axion contribution to the CXB is given by
$$
6.1 \cdot 10^{-12}~\hbox{erg cm}^{-2} \hbox{ s}^{-1} \hbox{deg}^{-2}  \left( \frac{B}{\sqrt{2} \hbox{ nG}} \frac{10^{13} \hbox{GeV}}{M} \right)^2,
$$
again showing that even for $M \gg 10^{11}~\hbox{GeV}$ is it easy to generate an observationally significant flux provided cosmological magnetic fields are close to the upper limit of $1 \hbox{ nG}$.

To conclude, primordial axions from modulus decay at $z \sim 10^{12}$ are a well motivated scenario of dark radiation which predicts  a present day Cosmic
Axion Background with energies between $0.1-1$ keV. Conversion of axions into photons  in astrophysical and cosmic magnetic fields makes the CAB manifest through quasi-thermal soft X-ray excesses. This mechanism may be responsible for  the soft X-ray excesses observed in galaxy clusters, and can generate a truly diffuse contribution to the $0.1-1$ keV cosmic soft X-ray background. Assuming reheating by decays of a Planck-coupled particle,
the detection of a Cosmic Axion Background at a scale $E_a$ would also imply the existence of a modulus with mass $m_{\Phi} \sim \left( \frac{T_{CMB}}{E_a} \right)^2 M_{Pl}$.

{\bf Acknowledgements:} We thank Konstantin Zioutas for informing us about the soft X-ray excess from galaxy clusters. We thank Jo Dunkley and Alexander Schekochihin for discussions.
We acknowledge funding from a Royal Society University Research Fellowship and the ERC Starting Grant `Supersymmetry Breaking in String Theory'.


\begin{thebibliography}{99}




  \bibitem{11032976}
  A.~G.~Riess, L.~Macri, S.~Casertano, H.~Lampeitl, H.~C.~Ferguson, A.~V.~Filippenko, S.~W.~Jha and W.~Li {\it et al.},
  Astrophys.\ J.\  {\bf 730} (2011) 119
   [Erratum-ibid.\  {\bf 732} (2011) 129]
  [arXiv:1103.2976 [astro-ph.CO]].

\bibitem{12125226}
  G.~Hinshaw, D.~Larson, E.~Komatsu, D.~N.~Spergel, C.~L.~Bennett, J.~Dunkley, M.~R.~Nolta and M.~Halpern {\it et al.},
  arXiv:1212.5226 [astro-ph.CO].

\bibitem{12126267}
  Z.~Hou, C.~L.~Reichardt, K.~T.~Story, B.~Follin, R.~Keisler, K.~A.~Aird, B.~A.~Benson and L.~E.~Bleem {\it et al.},
  arXiv:1212.6267 [astro-ph.CO].

\bibitem{13010824}
  J.~L.~Sievers, R.~A.~Hlozek, M.~R.~Nolta, V.~Acquaviva, G.~E.~Addison, P.~A.~R.~Ade, P.~Aguirre and M.~Amiri {\it et al.},
  arXiv:1301.0824 [astro-ph.CO].

\bibitem{13035076}
  P.~A.~R.~Ade {\it et al.}  [ Planck Collaboration],
  arXiv:1303.5076 [astro-ph.CO].

\bibitem{12083562}
  M.~Cicoli, J.~P.~Conlon and F.~Quevedo,
  Phys.\ Rev.\ D {\bf 87}, 043520 (2013)
  [arXiv:1208.3562 [hep-ph]].

\bibitem{12083563}
  T.~Higaki and F.~Takahashi,
  JHEP {\bf 1211}, 125 (2012)
  [arXiv:1208.3563 [hep-ph]].
\bibitem{13047987}
  T.~Higaki, K.~Nakayama and F.~Takahashi,
  arXiv:1304.7987 [hep-ph].

\bibitem{09063297}
  R.~Blumenhagen, J.~P.~Conlon, S.~Krippendorf, S.~Moster and F.~Quevedo,
  JHEP {\bf 0909} (2009) 007
  [arXiv:0906.3297 [hep-th]].

  \bibitem{hepph0503216}
  K.~Choi, A.~Falkowski, H.~P.~Nilles and M.~Olechowski,
  Nucl.\ Phys.\ B {\bf 718} (2005) 113
  [hep-th/0503216].

\bibitem{08040863}
  B.~S.~Acharya, P.~Kumar, K.~Bobkov, G.~Kane, J.~Shao, and S.~Watson,
  JHEP {\bf 0806} (2008) 064
  [arXiv:0804.0863 [hep-ph]].

\bibitem{13041804}
  J.~P.~Conlon and M.~C.~D.~Marsh,
  arXiv:1304.1804 [hep-ph].

\bibitem{Sikivie1}
  P.~Sikivie,
  Phys.\ Rev.\ Lett.\  {\bf 51}, 1415 (1983)
  [Erratum-ibid.\  {\bf 52}, 695 (1984)].
\bibitem{Sikivie2}
  P.~Sikivie,
  Phys.\ Rev.\ D {\bf 32}, 2988 (1985)
  [Erratum-ibid.\ D {\bf 36}, 974 (1987)].


\bibitem{Peccei:1977hh}
  R.~D.~Peccei and H.~R.~Quinn,
  Phys.\ Rev.\ Lett.\  {\bf 38}, 1440 (1977).



\bibitem{Raffelt:1987im}
  G.~Raffelt and L.~Stodolsky,
  Phys.\ Rev.\ D {\bf 37}, 1237 (1988).


\bibitem{Govoni:2004as}
  F.~Govoni and L.~Feretti,
  Int.\ J.\ Mod.\ Phys.\ D {\bf 13}, 1549 (2004)
  [astro-ph/0410182].


\bibitem{Durret:2008jn}
  F.~Durret, J.~S.~Kaastra, J.~Nevalainen, T.~Ohashi and N.~Werner, Space Science Reviews, {\bf 134}, 51 (2008)
  arXiv:0801.0977 [astro-ph].

\bibitem{Lieu1996}
R. Lieu et al, \emph{Science}:274, 1335,

\bibitem{Bonamente2002}
  M.~Bonamente, R.~Lieu, M.~K.~Joy and J.~H.~Nevalainen, Astrophys.\ J.\ {\bf 576}, 688 (2002),
  astro-ph/0205473.

\bibitem{Bonamente2001}
  M.~Bonamente, R.~Lieu, J.~Nevalainen and J.~S.~Kaastra, Astrophys.\ J.\ {\bf 552}, L7 (2002),
  astro-ph/0103331.

\bibitem{Nevalainen2003}
  J.~Nevalainen, R.~Lieu, M.~Bonamente and D.~Lumb,
  Astrophys.\ J.\  {\bf 584}, 716 (2003)
  [astro-ph/0210610].

  \bibitem{Henriksen2003}
  M.~J.~Henriksen, D.~S.~Hudson and E.~Tittley,
  Astrophys.\ J.\  {\bf 610}, 762 (2004)
  [astro-ph/0306341].

\bibitem{Bonamente2007}
  M.~Bonamente, J.~Nevalainen and R.~Lieu,
  Astrophys.\ J.\  {\bf 668}, 796 (2007)
  [arXiv:0707.0992 [astro-ph]].

\bibitem{Lehto2010}
  T.~Lehto, J.~Nevalainen, M.~Bonamente, N.~Ota and J.~Kaastra,
  Astronomy and Astrophysics, {\bf 524}:A70,
  arXiv:1010.3840 [astro-ph.CO].

  \bibitem{Bonamente:2009ns}
  M.~Bonamente, R.~Lieu and E.~Bulbul,
  Astrophys.\ J.\  {\bf 696}, 1886 (2009)
  [arXiv:0903.3067 [astro-ph.CO]].

\bibitem{10020594}
  A.~Bonafede, L.~Feretti, M.~Murgia, F.~Govoni, G.~Giovannini, D.~Dallacasa, K.~Dolag and G.~B.~Taylor,
  arXiv:1002.0594 [astro-ph.CO].

\bibitem{Ensslin1999}
T. A. Enßlin, R. Lieu, P. L. Biermann, Astronomy and Astrophysics, {\bf 344}:409 (1999)

\bibitem{Bowyer2004}
S. Bowyer, E. J. Korpela, M. Lampton, T. W. Jones, Astrophys.\ J.\ {\bf 605}, 168 (2004)

\bibitem{AtoyanVolk}
  A.~M.~Atoyan and H.~J.~Volk,
   Astrophys.\ J.\  {\bf 535}, 45 (2000)
  astro-ph/9912557.

  \bibitem{TsayBowyer}
  M.~Y.~Tsay, C.~-Y.~Hwang and S.~Bowyer,
  Astrophys.\ J.\  {\bf 566}, 794 (2002)
  astro-ph/0111072.

\bibitem{Han}
 J.~Han, C.~S.~Frenk, V.~R.~Eke, L.~Gao, S.~D.~M.~White, A.~Boyarsky, D.~Malyshev and O.~Ruchayskiy,
  Mon.\ Not.\ Roy.\ Astron.\ Soc.\  {\bf 427}, 1651 (2012)
  [arXiv:1207.6749 [astro-ph.CO]].


\bibitem{Ando}
  S.~Ando and D.~Nagai,
  JCAP {\bf 1207}, 017 (2012)
  [arXiv:1201.0753 [astro-ph.HE]].

\bibitem{Kim1989}
Kim, K.-T., Kronberg, P.~P., Giovannini, G., \& Venturi, T. 1989, Nature, 341, 720


\bibitem{09110015}
  A.~Mirizzi and D.~Montanino,
  JCAP {\bf 0912} (2009) 004
  [arXiv:0911.0015 [astro-ph.HE]].


  \bibitem{0512542}
  R.~C.~Hickox and M.~Markevitch,
  Astrophys.\ J.\  {\bf 645} (2006) 95
  [astro-ph/0512542].

\bibitem{0702556}
  R.~C.~Hickox and M.~Markevitch,
  Astrophys.\ J.\  {\bf 661} (2007) L117
  [astro-ph/0702556].

\bibitem{12084105}
  N.~Cappelluti, P.~Ranalli, M.~Roncarelli, P.~Arevalo, G.~Z.~A.~Comastri, R.~Gilli, E.~Rovilos and C.~Vignali {\it et al.},
  Mon.\ Not.\ Roy.\ Astron.\ Soc.\  {\bf 427}, 651 (2012)
  [arXiv:1208.4105 [astro-ph.CO]].









\end{thebibliography}
\end{document}